# Highly uniform GaSb quantum dots with indirect-direct bandgap crossover at telecom range

*Abhiroop Chellu, Joonas Hilska, Jussi-Pekka Penttinen, Teemu Hakkarainen\**

Optoelectronics Research Centre, Physics Unit, Tampere University, Korkeakoulunkatu 3, 33720 Tampere, Finland.

\*Corresponding author: teemu.hakkarainen@tuni.fi

**ABSTRACT**

We demonstrate a new quantum-confined semiconductor material based on GaSb quantum dots (QDs) embedded in single-crystalline AlGaSb matrix by filling droplet-etched nanoholes. The droplet-mediated growth mechanism allows formation of low QD densities required for non-classical single-QD light sources. The photoluminescence (PL) experiments reveal that the GaSb QDs have an indirect-direct bandgap crossover at telecom wavelengths. This is due to the alignment of the $\Gamma$ and $L$ valleys in the conduction band as a result of quantum confinement controlled by dimensions of the nanostructure. We show that in the direct bandgap regime close to 1.5 µm wavelength, the GaSb QDs have a type I band alignment and exhibit excitonic emission with narrow spectral lines and very low inhomogeneous broadening of PL emission owing to the high material quality and dimensional uniformity. These properties are extremely promising in terms of applications in infrared quantum optics and quantum photonic integration.

## I. INTRODUCTION

Semiconductor QDs embedded in a single-crystalline host matrix are important building blocks of emerging quantum technologies based on non-classical light sources, such as single- or entangled-photon emitters.[1-3] Such quantum nanostructures can be fabricated by several techniques including Stranski-Krastanov (SK) growth,[4,5] droplet epitaxy,[6-8] growth inside pyramidal holes,[9] vapor-liquid-solid growth in nanowires,[10] and filling of nanoholes formed by local droplet etching (LDE).[11] The approach based on filling nanoholes is particularly interesting due to its advantages which include narrow exciton linewidths,[12] extremely small inhomogeneous broadening due to size uniformity,[13] bright single-photon emission,[14] and vanishing fine structure splitting owing to the dimensional symmetry and lack of strain-induced piezoelectric asymmetries.[15] These properties have enabled the use of GaAs/AlGaAs QDs grown by filling nanoholes in non-classical light sources providing state-of-the-art performance in terms of photon indistinguishability and entanglement.[14,16-18] However, the operation of the light sources based on this approach is restricted to the 680-800 nm spectral range[13,19] due to the limited direct bandgap range of (Al)GaAs alloys. While increasing efforts have been dedicated to develop non-classical light sources based on QDs emitting in the telecom O and C-bands using both droplet epitaxy[20,21] and strain-induced growth,[21-24] their performance in terms of photon indistinguishability and fidelity of entanglement is still lacking behind the best demonstrations based on LDE GaAs/AlGaAs QDs.[18,25]

The reports on LDE of nanoholes towards QD formation were limited to the (Al)GaAs alloy system[11,26-33] until recently when we demonstrated highly controllable LDE of AlGaSb with Ga droplets[34]. In contrast to the LDE of arsenides, the low vapor pressure of Sb allows deterministic control of the group V beam pressure which essentially controls nanohole etching process by adjusting the Sb source needle valve. Furthermore, the GaSb-based materials provide suitable bandgaps for accessing the technologically important telecom wavelengths[35] and even beyond that to the mid-infrared wavelength range[36] with compatibility for silicon photonic integration.[37] The additional benefits of the AlGa(As)Sb alloys include a very high refractive index contrast[38] (exceeding that of AlGaAs), which is important for constructing photonic devices. Moreover, the lattice mismatch between GaSb-based materials and dissimilar substrates can be relaxed right at the first interface by formation of a network of 90°-dislocations[39,40] and exploitation of nucleation layers,[41] which is particularly useful considering direct growth of QD emitters on silicon waveguides for chip-level quantum photonic integration.

Here, we report a new quantum-confined material based on GaSb QDs in AlGaSb matrix formed by filling droplet-etched nanoholes. We use the same methodology as in our previous work on nanohole etching[34] with the exception that Al droplets are used instead of Ga in order to avoid charge carrier confinement in the ring structure formed around the nanoholes in the LDE process. Optically active and highly uniform QDs are then formed by depositing a thin GaSb quantum well (QW) on the droplet-etched surface, similarly to the GaAs/AlGaAs system.[11] The resulting QDs exhibit excitonic emission at telecom wavelengths with type I band alignment contrary to the existing antimonide QDs such as InSb/InAs,[42] InSb/GaAs, and GaSb/GaAs QDs,[43,44] which confine only holes while the electrons are bound outside the QD.

## II. RESULTS AND DISCUSSION

Formation of GaSb QDs in an AlGaSb matrix begins with etching of nanoholes in AlGaSb surface with Al droplets using the growth process described in Section IV. Figure 1 shows atomic force microscopy (AFM) images of the AlGaSb surfaces after droplet etching at 500°C and 395°C with Al droplets formed by deposition of different Al coverages $\theta_{Al}$. As shown in Fig. 1(a), LDE at 500°C with $\theta_{Al}$=3.2 monolayers (ML) results in mobile droplets and inhomogeneous nanoholes. For $\theta_{Al}$=2.0 ML, we observed well-defined nanoholes, but their size distribution is bimodal, as shown in Fig. 1(b). Similar morphologies have been observed in AlGaAs surfaces after LDE with Al droplets due to a change in surface reconstruction at high Al coverages.[45,46] For $\theta_{Al}$ ranging from 1.5 ML to 1.25 ML, we observe a monomodal distribution of uniform nanoholes. No nanoholes are observed for $\theta_{Al}$=1.15 ML (See Fig. S1 in the supplementary material), indicating that the critical coverage for Al droplet formation on AlGaSb is approximately 1.2 ML which is consistent with our findings on Ga droplet formation on the same surface,[34] as well as with the group III metal coverage required for saturating the (1x3) surface reconstruction.[47] In case of LDE at 395°C, a bimodal nanohole distribution with additional mobile droplets is observed for $\theta_{Al}$=2.0 ML (Fig. 1(e)). A monomodal distribution of uniform nanoholes is again recovered by reducing $\theta_{Al}$ down to 1.5 ML. The densities of the uniform nanoholes formed at 500°C and 395°C are $1.9\times10^6$ cm$^{-2}$ and $2.6\times10^7$ cm$^{-2}$ respectively, both of which are low enough to allow addressing individual QDs in optical experiments. According to the cross-sectional profiles shown in Figs. 1(g) and 1(f), the nanohole formed at 500°C with $\theta_{Al}$=1.5 ML is 70 nm deep, while the nanohole formed at 395°C with the same coverage is around 12 nm deep. This reduction in nanohole depth is a result of the Al droplets becoming smaller and denser as the temperature is decreased. Consequently, the lateral dimensions of the droplet-etched nanohole also decrease as the temperature is reduced. The slope of the nanohole sidewalls is around 55° which matches with the (111) crystal planes. It should be noted that the temperatures used in this work for LDE of AlGaSb with Al droplets are significantly lower than the bulk melting point of Al which is 660°C. Successful droplet-etching at such low temperatures can be in part explained by the melting point depression in nanoscale droplets. Furthermore, it has been previously shown for LDE of AlGaAs with Al droplets that dissolution of Ga atoms from the surface into the Al droplet can lower the melting point and enable droplet etching at temperatures as low as 360°C.[48]

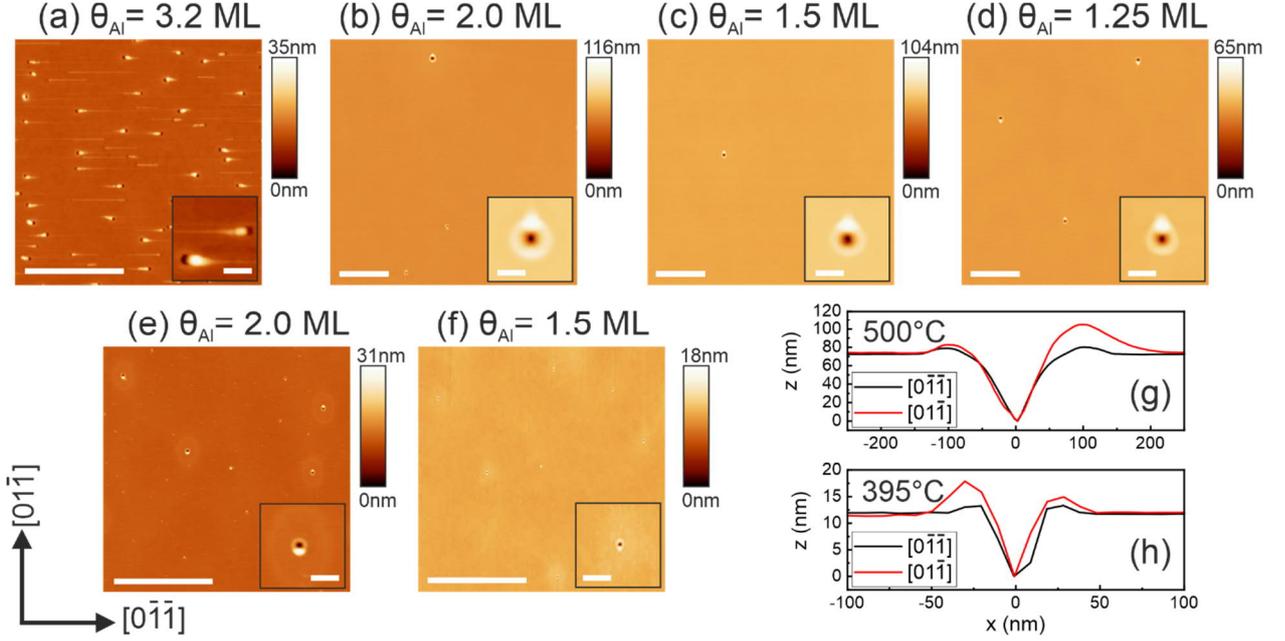

**FIG. 1.** AFM images of the AlGaSb surface after nanohole etching with Al droplets. (a)-(d) show the nanoholes formed at 500°C and (e)-(f) at 395°C by deposition of different amounts of Al as measured by the monolayer coverage $\theta_{Al}$. The cross-sectional profiles of nanoholes etched with $\theta_{Al}$=1.5 ML in (g) and (h) are taken from (c) and (f), respectively, for both $[0\bar{1}1]$ and $[0\bar{1}\bar{1}]$ crystal directions. The scale bars are 2 μm in the main images and 200 nm in the insets in (a)-(f).

Optically active GaSb nanostructures are then formed by filling the 12 nm deep nanoholes shown in Fig. 1(f) by depositing GaSb QWs with different thicknesses and embedding them in AlGaSb. The room temperature PL of such structures is shown in Fig. 2. As shown in Fig. 2(a), for QW thicknesses of 0 nm and 1.2 nm, we observe the PL only from the AlGaSb matrix at around 1120 nm, while no PL emission at all is detected in case of 2.4 nm. For QW thicknesses from 3.7-6.8 nm we observe the GaSb QW peak, which redshifts and intensifies as the QW thickness is increased, as summarized in Fig. 2(b). These results can be explained by looking at the GaSb bandstructure which is illustrated in Fig. 2(c). While GaSb in bulk form is a direct bandgap semiconductor with conduction band minimum at $\Gamma$ point, the energy separation between the $\Gamma$ and $L$ points is relatively small. The electron effective mass at $\Gamma$ is $m^*(\Gamma) = 0.039 m_0$, where $m_0$ is the electron mass, while at the asymmetric $L$ valley we have longitudinal and transversal components $m_l^*(L) = 1.3 m_0$ and $m_t^*(L) = 0.1 m_0$.[36] Because of this significant difference in the effective masses, the quantum effect is more pronounced for the $\Gamma$ electrons than for the $L$ electrons which results in direct-indirect bandgap crossover as the dimensions of the nanostructure are reduced, as reported for GaSb/AlSb QWs.[49] In order to obtain a quantitative description for the crossover, we solve the electron energy states in the GaSb/AlGaSb QW using the effective mass approach with finite barriers. In case of confinement in the [100] direction (the growth direction) we use a projected effective mass of $0.52 m_0$ for the $L$ electrons,[49] while for the confinement in the [111] direction, the longitudinal mass $m_l^*(L)$ can be used. The band nonparabolicities are accounted as in[49] by corrections to the effective masses as $m_{NP}^*(\Gamma) = m^*(\Gamma)\left(1 + \frac{2.5\text{ nm}}{t_{QW}}\right)$ for $\Gamma$ and $m_{NP}^*(L) = m^*(L)\Big(1 +$ $\frac{0.6 \text{ nm}}{t_{QW}}$) for $L$, with $t_{QW}$ being the QW thickness in nanometers. The solutions for the lowest electron states for $\Gamma$ and $L$ points are shown in Fig. 2(d), which confirms that the energy structure of the QW is indirect when its dimensions are small. In case of the [100] confinement, the crossover from indirect to direct bandgap is observed when the QW thickness is increased beyond approximately 6 nm, which is consistent with the room-temperature PL intensity trend shown in Fig. 2(b). In case of [111] confinement, the crossover is observed at slightly larger QW thickness due to the larger effective mass. It should be noted, however, that the precise position of the crossover point is sensitive to the bulk value used for the $\Gamma$-$L$ energy separation. We have used a value of $E_L$-$E_\Gamma$=0.063 eV as recommended by Vurgaftman *et al.*[36] However, the values reported in the literature range from 0.063 meV to 0.1 meV,[50-52] and use of a larger energy separation would lead to a crossover at slightly smaller $t_{QW}$.

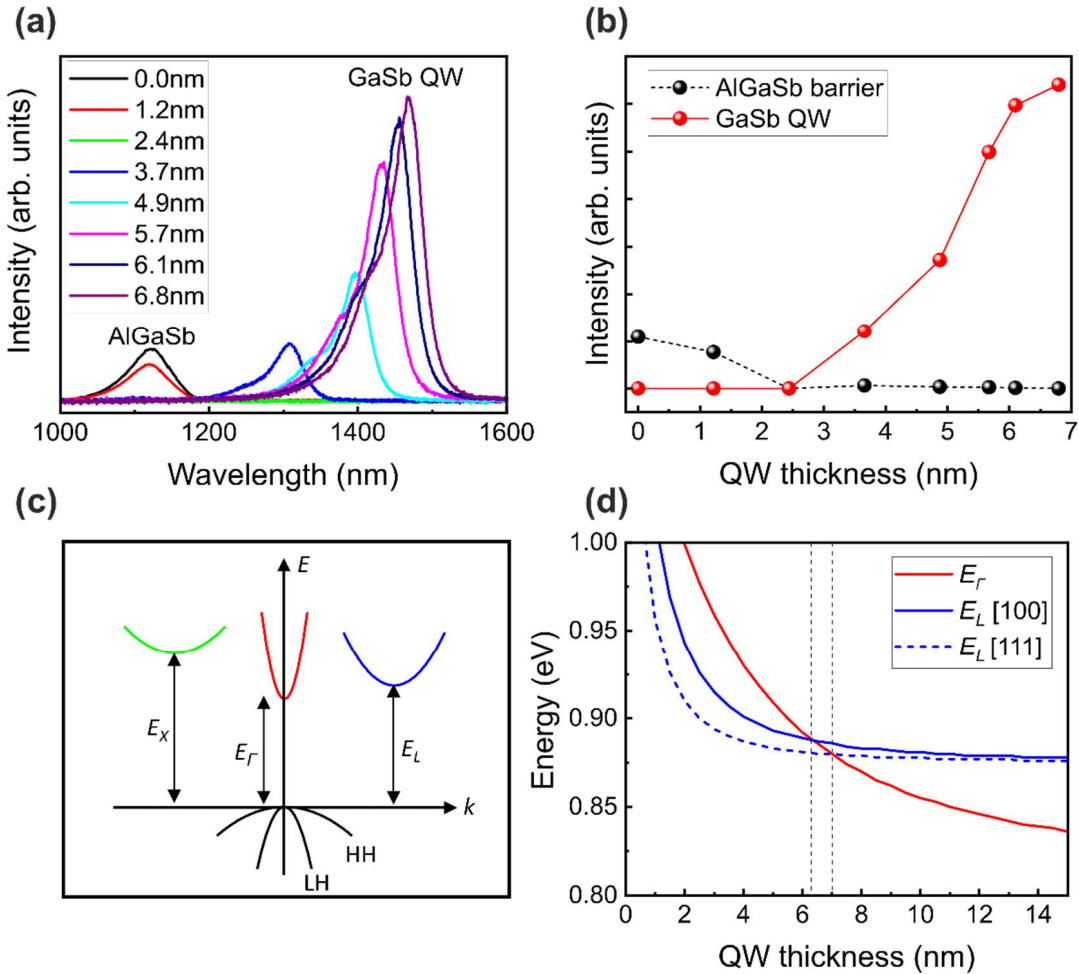

**FIG. 2.** (a) Room-temperature PL spectra showing the AlGaSb barrier and GaSb QW emission for different QW thicknesses. (b) The intensities of the peaks from (a) as a function of the QW thickness. (c) A simplified energy band diagram of GaSb bulk including heavy hole (HH) and light hole (LH) valence bands and $\Gamma$, $L$, and $X$ valleys in the conduction band. (d) Energies of the lowest quantum confined electron states in $\Gamma$ and $L$ valleys as a function of QW thickness. The energies are referenced with respect to the top of the valence band. The vertical dashed lines indicate the indirect-direct bandgap crossover for confinements in the [100] and [111] directions.

Similar trends in the GaSb QW PL peak intensity and position with increasing $t_{QW}$ are observed also at 20 K (Fig. 3(a)). Due to the absence of processes involving acoustic phonons, the QW

PL intensity change around the crossover point is more rapid at low temperature than at room temperature. Consequently, we observe a 50-fold increase in QW peak intensity when $t_{QW}$ is increased from 3.7 nm to 4.9 nm, which is consistent with the low-temperature PL reported for GaSb/AlSb QWs.[49,53] The full trend of QW PL intensity as a function of $t_{QW}$ is presented in Fig. S3 in the supplementary material. The low-temperature PL spectra in Fig. 3(a) also reveal emission from the GaSb substrate. It should be noted that due to the rapid extinction of the 532 nm excitation laser in the AlGaSb material, the GaSb substrate luminescence is actually excited by the backward emitted QW luminescence instead of by the laser light directly. Consequently, we observe a direct proportionality between the GaSb substrate and QW PL intensities (Fig. S3). In addition to the GaSb QW and substrate peaks, the PL spectra of the samples with 5.7 nm and 6.1 nm QWs grown on the droplet-etched surface exhibit additional narrow peaks P1-P3 which are not present in the reference QW grown on a planar AlGaSb surface without nanoholes. These emission peaks can be attributed to the localized emission at the nanoholes. As revealed by the PL spectra obtained with different excitation power densities (Fig. 3(b)) and the power-dependency of the peak intensities (Fig. 3(c)), the P1 and P2 peaks exhibit a typical behavior of QD ground state and first excited state emission, respectively. The intensity $I$ of the P1 peak exhibits a linear dependency of the PL intensity on the excitation power $P$, i.e. $I \sim P^{1.00}$, which is a sign of typical excitonic behavior at the ground state of the localized system. Peak P2, on the other hand, shows superlinear power-dependency with $I \sim P^{1.37}$ and emerges at higher excitation powers than P1 does, which is typical behavior of the first excited state.[54] The intensity of P2 also keeps increasing as a function of excitation power beyond the point of saturation of P1 as expected for the excited state emission. The P1-P2 peak separation in Fig. 3(b) is 37 meV, which is similar to the energy level spacing observed in GaAs/AlGaAs QDs.[11] P3 peak shows a linear power-dependency and emerges at lower excitation powers than P2, which suggest that it is a ground state of another localized system rather than the second excited state of the QD inside the nanohole. A similar PL peak has been observed in case of GaAs/AlGaAs QDs grown by filling droplet-etched nanoholes and attributed to localized QW emission at the edge of ring structure outside the nanohole.[55] Since the PL signal was collected from a large number of QDs, the width of the P1 peak below saturation level represents the inhomogeneous broadening as a result of QD size dispersion. As shown in the inset in Fig.3(b), the full-width-at-half-maximum (FWHM) for the P1 peak is just 8 meV, which indicates significantly smaller inhomogeneous broadening than in InAs SK QDs (typically tens of meV)[56] and comparable to the GaAs/AlGaAs QDs formed by filling droplet-etched nanoholes.[11,13]

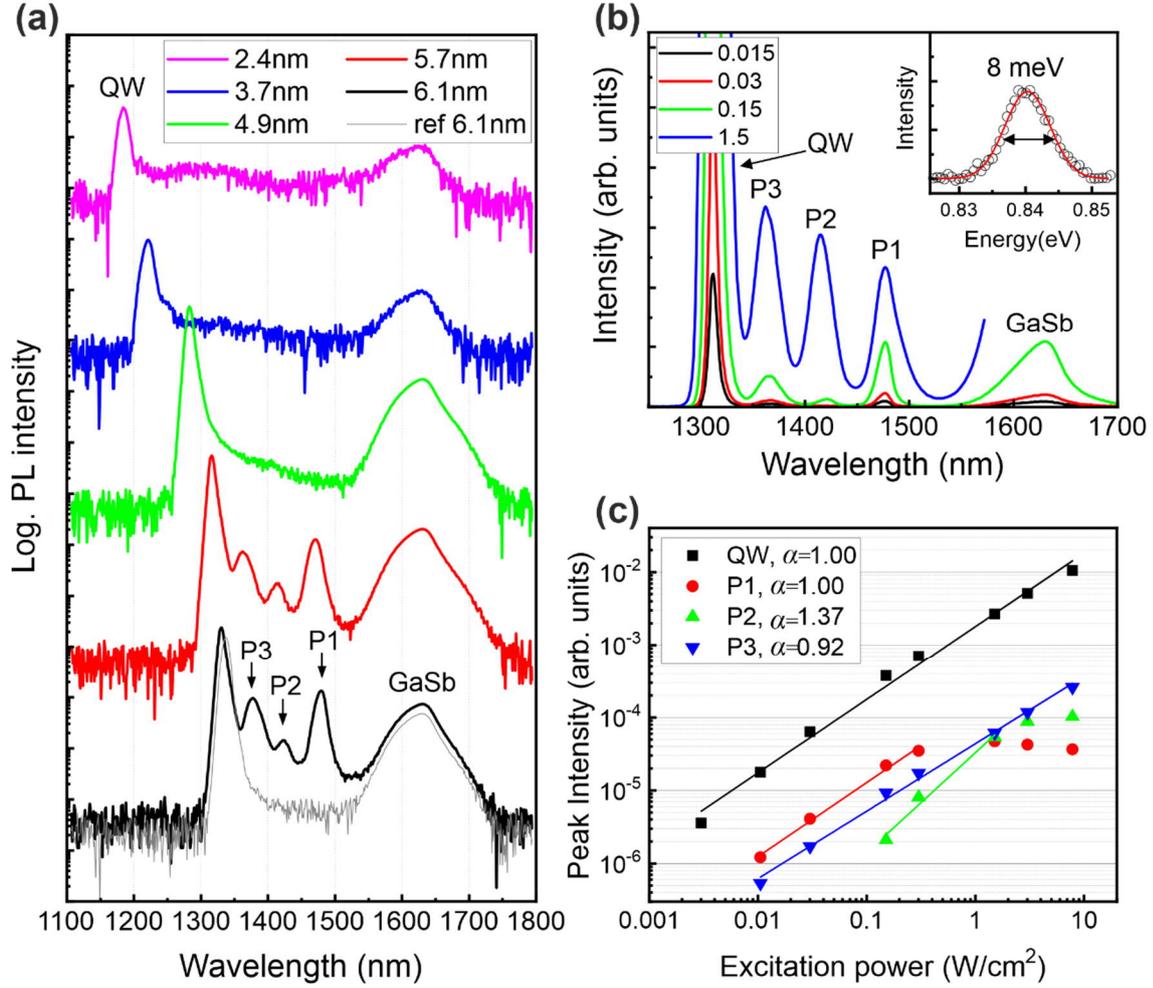

**FIG. 3.** (a) PL spectra at 20 K showing the GaSb QW and GaSb bulk emission peaks as well as the QD peaks P1-P3 arising from emission in a large number of filled nanoholes. The reference spectrum presented by the gray line is measured from a 6.1 nm thick GaSb QW grown on a planar AlGaSb surface without nanoholes. (b) Excitation power-dependent spectra for the 5.7 nm sample, with laser power density indicated in W/cm$^2$. The inset shows the P1 peak width for the lowest excitation power. The intensities ($I$) of the QW and QD peaks as a function of excitation power ($P$) are shown in (c). The solid lines represent fits with $I = A \times P^\alpha$, where $A$ is a constant and $\alpha$ is the exponent depending on the decay mechanism.

It should be noted that no QD emission is observed for QW thicknesses below 5.7 nm, and then from 5.7 nm to 6.1 nm we observe a red shift of the P1 peak from 1470 nm to 1478 nm similarly to the QW emission in the direct bandgap regime. This is due to the crossover from indirect to direct bandgap as the size of the nanostructure increases, just like shown for the QW in Fig. 2. Since the QDs were formed by completely filling 12 nm deep nanoholes (See Fig. S2 in the supplementary material), the critical dimensions for the indirect-direct crossover appear to be larger for GaSb QDs than for GaSb QWs. The quantum effect can be expected to be larger in a zero-dimensional structure, but here, the nanohole morphology brings additional confinement specifically in the [111] directions for which the electrons in the *L* valley experience the large longitudinal effective mass of 1.3$m_0$. In such case, the critical size required for the indirect-direct crossover is expected to be larger than for the confinement in the [100] direction as shown in Fig. 2(d).

The luminescence properties of the GaSb/AlGaSb QDs formed by filling droplet-etched nanoholes were further investigated by micro-PL spectroscopy around the P1 peak wavelength. As shown in multi-QD spectra (Fig. 4(a)), the QDs exhibit emission characterized by narrow exciton lines. The excitation power-dependent single-QD measurement reveals a dominant single-exciton (X) peak which follows the power law dependency $I \sim P^{\alpha}$ with $\alpha=1.1$, as shown in Fig. 4(b) and (c). The width of the X line in Fig. 4(b) is 90 µeV, which is at the resolution limit of our optical setup. These results prove that the GaSb/AlGaSb QDs possess typical luminescence characteristics of excitonic recombination in a high-quality, 0-dimensional quantum system with type I band alignment. This is in contrast to other antimonide QDs, which have type II systems and thus exhibit broad emission peaks and blueshift of the exciton lines with increasing excitation power.[57] No biexciton emission is identified in the single-QD spectra. Instead, several peaks with $\alpha<2$ appear at lower energies when the excitation power is increased. These additional lines are most probably a collection of charged and neutral multi-exitonic lines as well as emission from other QDs. Such PL characteristics are typically observed for GaAs/AlGaAs QDs grown by filling droplet-etched nanoholes, with a clear biexciton emission being detectable using resonant excitation.[16]

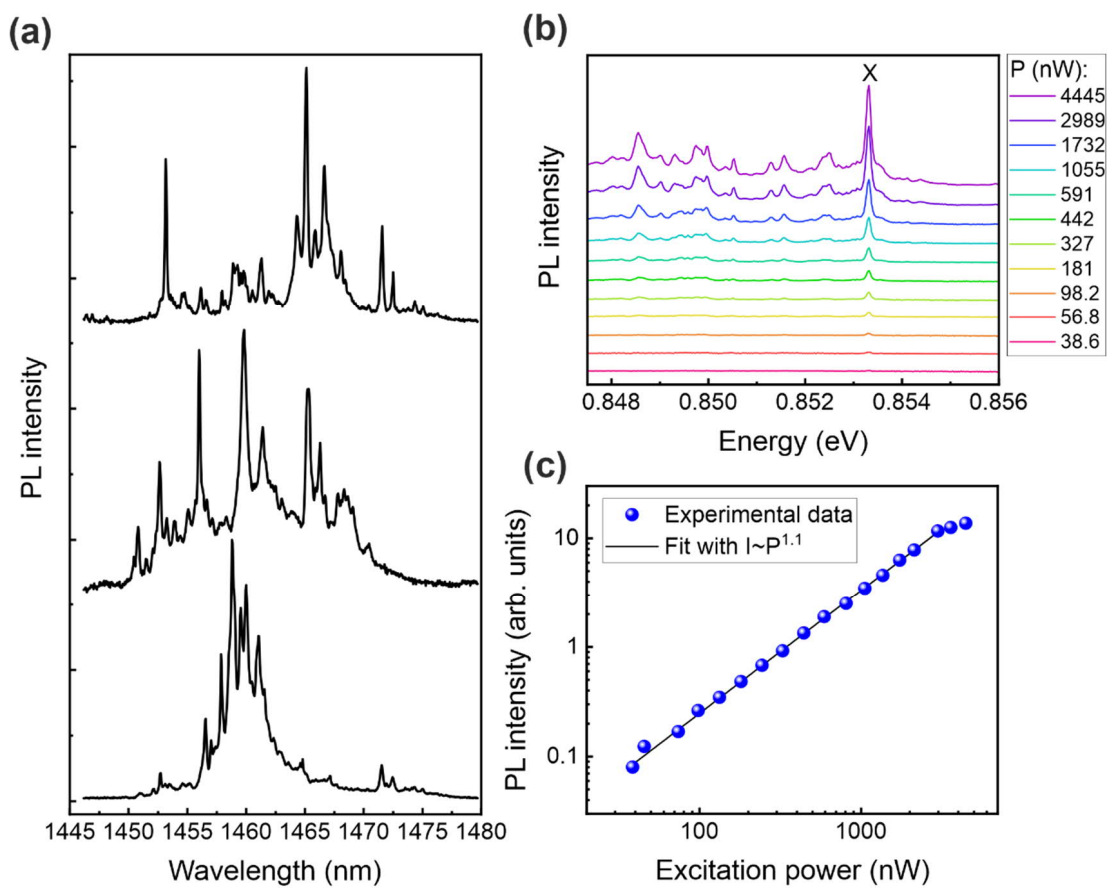

**FIG. 4.** Micro-PL results obtained at 6 K for the sample with a 5.7 nm thick GaSb layer deposited on droplet-etched AlGaSb surface. (a) Emission from multiple QDs collected from three different locations. (b) Single-QD emission measured with different excitation powers. (c) The intensity of the single-exciton (X) peak from (b) plotted as a function of excitation power.

The PL observations show the potential of the GaSb/AlGaSb QDs for non-classical light emission at infrared and telecom wavelengths. An additional benefit of the LDE-based QD formation mechanism is that the band structure of the QD material is widely tunable by alloying GaSb with other elements, such as As, In, and Bi. This provides the freedom to extend the emission wavelength to even longer wavelengths towards 2 µm and potentially beyond, and to simultaneously adjust the alignment of the $\Gamma$ and $L$ valleys in order to control the charge carrier recombination processes. Furthermore, the indirect-direct bandgap crossover could be utilized for triggering spontaneous emission in the QDs by interfacing them with piezoelectric strain elements or surface acoustic waves.[58]

## III. CONCLUSION

In conclusion, we have demonstrated a new type I semiconductor quantum system based on highly uniform GaSb QDs formed in AlGaSb matrix by filling droplet-etched nanoholes. This technique allows the formation of low QD densities required for single-QD quantum light sources. The GaSb QDs exhibit an indirect-direct bandgap crossover at the telecom wavelength range, and thus allow access to a quantum system which can be tuned from highly luminescent to non-luminescent in a controllable way. In the direct bandgap regime at around 1.470 µm wavelength, the GaSb QDs exhibit PL emission with narrow exciton lines and very small inhomogeneous broadening which are extremely promising findings in terms of applications in infrared quantum optics and quantum photonic integration.

## IV. EXPERIMENTAL SECTION

### A. Sample growth and layer structures

All samples were grown by solid-source molecular beam epitaxy (MBE) on n-GaSb(100) substrates. For the determination of nanohole density and morphology, we prepared samples following the procedure as described in.[34] Namely, after growth of a GaSb buffer layer and a 100 nm thick $Al_{0.3}Ga_{0.7}Sb$ layer at 500°C, the substrate was set to pyrometer temperatures of 395°C or 500°C for the droplet deposition and LDE steps. Al was deposited for different target coverages between $\theta_{Al}$=1.15-3.2 ML under a predetermined Sb-flux of 0.067 ML/s. The nanoholes were formed by letting the Al droplets react with the $Al_{0.3}Ga_{0.7}Sb$ surface under the same Sb-flux for 180 s, which was a long enough annealing time to ensure complete etching of the nanoholes and consumption of the liquid droplet.

For optical studies, the GaSb-filled nanoholes were grown on a 100 nm thick AlGaSb layer by following the same steps as above, with the GaSb QD filling performed immediately after the LDE step at the LDE temperature. After filling the nanoholes, the sample temperature was increased to 500°C while simultaneously capping the GaSb QDs with 100 nm of $Al_{0.3}Ga_{0.7}Sb$. To promote recombination in the QW/QD structure, the AlGaSb films were cladded on either side with 50 nm thick lattice-matched AlAsSb layers which confined the generated carriers to the AlGaSb matrix. Finally, to prevent oxidative degradation of the topmost AlAsSb layer under atmosphere, the structure was capped with 15 nm of GaSb. For all samples, the Ga and Al fluxes were 0.7 ML/s and 0.3 ML/s, respectively. The full layer structures are illustrated in Fig. S4 in the supplementary material.

### B. Photoluminescence spectroscopy

Room temperature PL spectra of the GaSb QWs were recorded using 532 nm excitation. The PL emission was dispersed with a grating and collected with an InGaAs array detector.

The low temperature PL spectra of the GaSb QWs and QD ensembles were measured at 20 K in a closed-loop He-cooled cryostat. The samples were excited with a 532 nm laser with a broad excitation spot having a diameter of 1.3 mm FWHM. The PL emission was collected and focused with lenses towards a 0.5 m monochromator and then detected with an InGaAs photodiode. Standard lock-in technique with thermoelectric cooling of the detector was used for improving the signal-to-noise ratio.

Single-QD emission was studied by micro-PL at 6 K using a low-vibration closed-loop He-cooled cryostat. The QDs were excited non-resonantly with a 965 nm laser diode through a confocal lens (f=2.75 mm, NA=0.68) and a 2 mm diameter hemispherical solid immersion lens. The PL emission from the QD was collected with the same lenses, dispersed with a 750 mm spectrograph, and detected with thermoelectric-cooled InGaAs array. The excitation spot size in the micro-PL experiments was around 2 µm excluding the effect of lateral diffusion of the photo-generated charge carriers.

## SUPPLEMENTARY MATERIAL

See supplementary material for additional AFM and photoluminescence data, and a schematic of the MBE-grown layer structures.

## ACKNOWLEDGMENT

The authors acknowledge financial support from the Academy of Finland Projects QuantSi (decision No. 323989) and NanoLight (decision No. 310985)

## DATA AVAILABILITY

The data that support the findings of this study are available from the corresponding author upon reasonable request.